\documentstyle[aps,epsf,multicol,prb]{revtex}
\input{psfig.sty}

\newcommand{\eps}{\varepsilon}

\begin{document}
\draft
\author{H. B. Weber$^{1\dag}$, 
R. H\"aussler$^1$, H. v. L\"ohneysen$^{1,2}$}
\address{$^{1}$ Physikalisches Institut, Universit\"at Karlsruhe, D-76128 
Karlsruhe}
\address{$^{2}$ Forschungszentrum Karlsruhe, Institut f\"ur
Nanotechnologie, D-76021 Karlsruhe}
\author{J. Kroha}
\address{Institut f\"ur Theorie der Kondensierten Materie, 
Universit\"at Karlsruhe, D-76128 Karlsruhe}
\date{\today}
\title{Non-equilibrium electronic transport and interaction in 
short metallic nanobridges}
\maketitle
\begin{abstract}
We have observed interaction effects in the differential conductance $G$
of short, disordered metal bridges in a well-controlled 
non-equilibrium situation, where the distribution function has a double 
Fermi step. 
A logarithmic scaling law is found both for the temperature 
and for the voltage dependence of $G$ in all 
samples. The absence of magnetic field dependence and the 
low dimensionality of our samples allow us to distinguish between
several possible interaction effects, proposed recently in nanoscopic samples. 
The universal scaling curve is explained quantitatively 
by the theory of electron-electron 
interaction in diffusive metals, adapted to the present case, where the
sample size is smaller than the thermal diffusion length. 
\end{abstract}

\begin{multicols}{2}
\narrowtext
\section{Introduction}
Dynamically screened electron-electron interactions  
in disordered metals are known to 
cause singular corrections to the electronic density of states (DOS),
as explained by Aronov and Al'tshuler  \cite{AA.85,altshuler.79}.
The corresponding zero-bias anomalies (ZBA) have been studied 
by tunneling spectroscopy on wide junctions in thermodynamic 
equilibrium \cite{Imry,Hertel}. How is the ZBA modified 
when a large current is driven through the system 
by a finite bias voltage, and the bridge is so small that there is phase 
coherence across the whole sample? 
In this case, no local equilibrium is reached, since 
energy relaxation does not occur within the sample. Instead, the distribution
function of electrons traversing the system exhibits two Fermi steps
related to the two different electrochemical potentials of the two leads. 
This fact has been claimed theoretically \cite{kulik,nagaev} and 
has recently been confirmed in tunneling experiments \cite{pothier.97} 
(where in addition the steps were rounded due to interactions in long wires) 
and indirectly by noise measurements in a similar setup \cite{strunk}. 

We have fabricated metallic nanobridges, which allow for the first
time to study the Aronov-Al'tshuler anomaly \cite{AA.85} in the stationary non-equilibrium 
situation specified above. A two-dimensional (2D),
diffusive metal bridge, much shorter than all inelastic scattering lengths,
is placed in good metallic contact between two thick reservoirs.
A high lead-to-bridge aspect ratio allows to maintain a finite voltage drop
across the bridge despite its small resistance, thus allowing the
observation of temperature-voltage scaling behavior. The study of a 
2D bridge is further motivated by the fact that this geometry allows
to distinguish between the Aronov-Al'tshuler conductance
anomaly and a possible two-channel Kondo 
effect induced by two-level systems \cite{zawa.98}, 
which has been put forward as the origin of ZBAs
observed in ultrasmall point contacts \cite{ralph.94}: 
In 3D both the Aronov-Al'tshuler and the two-channel Kondo anomalies   
show square root power-law behavior; in 2D the Aronov-Al'tshuler correction 
is logarithmic, while the two-channel Kondo singularity, as a local effect, 
is independent of dimension.
As will be discussed, the conductance through the non-equilibrium 
bridge can be well described by a two-termnal Landauer-B\"uttiker 
formula, generalized for interacting systems \cite{wingreen.92}. 
Applying the theory of electron-electron interaction in disordered systems in non-equilibrium,
we find that the electronic DOS is strongly modified by the 
non-equilibrium and displays two logarithmic singularities, 
corresponding to the two steps in the distribution function.
This yields quantitative agreement with the experimental results. 

\begin{figure} \epsfxsize7.7cm \centering\leavevmode\epsfbox{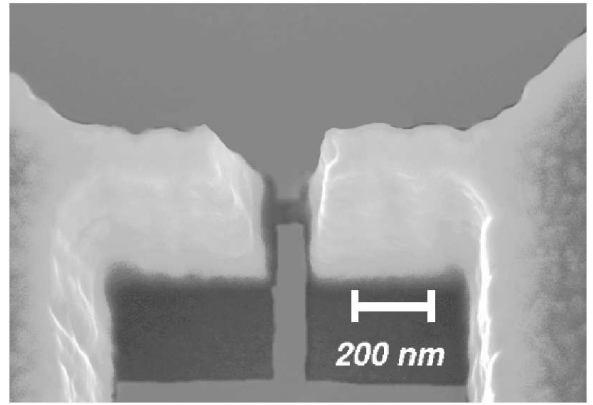}  
\vspace*{0.5cm} 	\caption{SEM picture of one of the samples, which were produced by the shadow-evaporation technique. The 
thin Cu$_{82}$Au$_{18}$ bridge (dark) is embedded  between two considerably thicker Cu leads. 
Dark regions are the thin shadow-evaporation replicas of the thick leads.
A thick replica of the bridge, which is not connected to the electrodes, 
is removed from the
picture for clarity.} 	\label{weber_fig1} \end{figure} \noindent

\section{Experiment}
The samples were produced by the shadow-evaporation technique \cite{Dunsmuir}. 
Metal was deposited at two different angles under UHV conditions through a 
PMMA mask patterned with e-beam lithography. We deposited 10\,nm of 
Cu$_{100-x}$Au$_{x}$ for the bridge at one angle (-13$^\circ$ off the 
normal direction of the surface) and 700 nm of Cu for the reservoirs 
(+13$^\circ$) at the opposite angle without breaking the vacuum (on the 
10$^{-9}$~mbar scale during evaporation).
The transport measurements were performed in a $^3$He/$^4$He dilution 
refrigerator in a temperature range between $T=100$~mK and $T=2.1$~K.
The differential resistance was measured with a LR~700 resistance 
bridge, which was capacitively coupled to a DC circuit through the 
sample. The AC excitation was less than $5~\mu$V.
\begin{minipage}{\linewidth}
\begin{table}
\begin{tabular}
{c|l|c c c c c c}
No. & & $d$  & $R$ & $D$  &$\ell$  & $A_{B=0\rm T}$ 
& $A_{B=8.5\rm T}$ \\
&&[nm]&[$\Omega$]&[cm$^2$/s]&[nm]& [$e^2/h$] & [$e^2/h$]\\
\hline
1 & Cu$_{82}$Au$_{18}$ & 10 & $13.6$ &34  &6.5 & $0.49$  &$0.48$ \\
2 & Cu$_{82}$Au$_{18}$ & 10 & $16$   &31  &6.0 & $0.51$  & --    \\
3 & Cu$_{50}$Au$_{50}$ & 20 & $6.5$  &37  &7.0 & $0.42$  & $0.45$\\
4 & Cu                 & 10 & $5.5$  &68  &13  & $0.7$   & $0.43$
\end{tabular} 
\vspace{0.2cm}
\caption{
Characteristic parameters for the four investigated nanobridges: 
material composition,  
thickness $d$, resistance $R$, diffusion coefficient $D$, 
mean free path $\ell$, 
and the amplitude $A$ 
of the zero-bias anomaly at an applied magnetic field
of $B=0$ and $B=8.5~$T, respectively (see text). 
In high magnetic field the pure Cu sample exhibits the same ZBA 
as all other samples, but in low field shows an enhanced amplitude of the
effect, while the scaling behavior Eqs.~(1), (2) remains valid. 
This needs further investigation.
}
\end{table}
\end{minipage}
The parameters characterizing the four different 
samples investigated in this study are summarized in table I.
Each of the four nanobridges is $L=80$~nm long, about 80~nm wide and their 
thickness $d$ is substantially smaller than the lateral extension $L$.
It is placed in good metallic contact between two bulk Cu leads, 
which are about 70 times thicker than the bridge and extend over a large area
of about 1~mm$^2$ each (see Fig.\,\ref{weber_fig1}). 
Hence the voltage applied to the sample drops only along the bridge, 
and the arising heating power is reliably conducted away by the leads. 
This was verified in a control experiment, where the cooling capability of 
the reservoirs was varied using a different geometry. 
No change of the behavior reported below was observed.
Moreover, as demonstrated in detail below, the electronic motion is
coherent over the entire size of the nanobridge, which means, in particular,
that there is no heat deposition within the bridge. 
The elastic mean free path $\ell$  was estimated from the 
residual-resistance ratio and consistently from the Drude formula. 
In all samples $\ell$ is 
comparable to the thickness $d$, but much shorter than the lateral 
length $L$, and therefore, the electronic density modes in the bridge 
obey the rules of  2D diffusive motion. 
Below we present experimental conductance data of sample no. (1); 
the results obtained  
from the other Cu$_{100-x}$Au$_x$ and Cu films are very similar.

\begin{figure}
\epsfxsize7.7cm
\centering\leavevmode\epsfbox{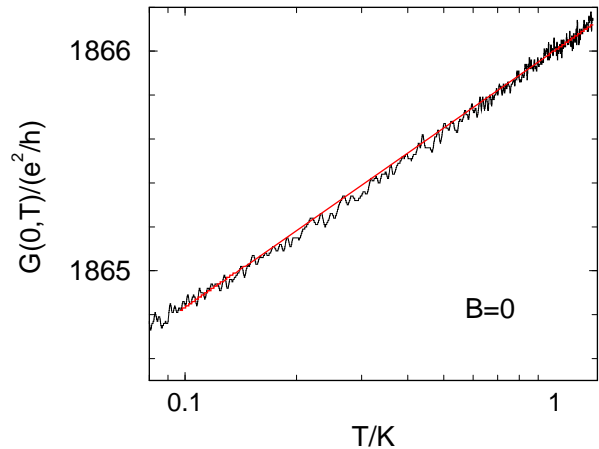}
  \vspace*{0.3cm}
	\caption{Zero-bias conductance of the Cu$_{82}$Au$_{18}$ bridge. 
The straight line is a fit according to Eq.~(\ref{scale1}).}
	\label{weber_fig2}
\end{figure}

Raw data of the zero-bias conductance as a function of temperature $T$ are 
shown in Fig.~\ref{weber_fig2}. We observe a logarithmic $T$ dependence between $T=100$~mK and $T=2$~K:
\begin{equation}
G(0,T)=G(0,T_o=1\rm{K})+\it A \cdot {\rm ln}(T/T_o)\; ,
\label{scale1}
\end{equation}
where $A$ denotes the amplitude of the effect, with $A = 0.49~e^2/h$
for this particular bridge. Below $T=100$~mK, small deviations from the
form Eq. (1) were observed, which are probably due to incomplete 
thermalization of the sample at the lowest temperatures.  
Such logarithmic behavior has often been observed in thin metallic films 
and metallic nanobridges and is usually attributed to weak localization  
\cite{schmid} and/or electron-electron interaction \cite{AA.85}. 

\begin{figure}[htb]
\epsfxsize8.1cm
\centering\leavevmode\epsfbox{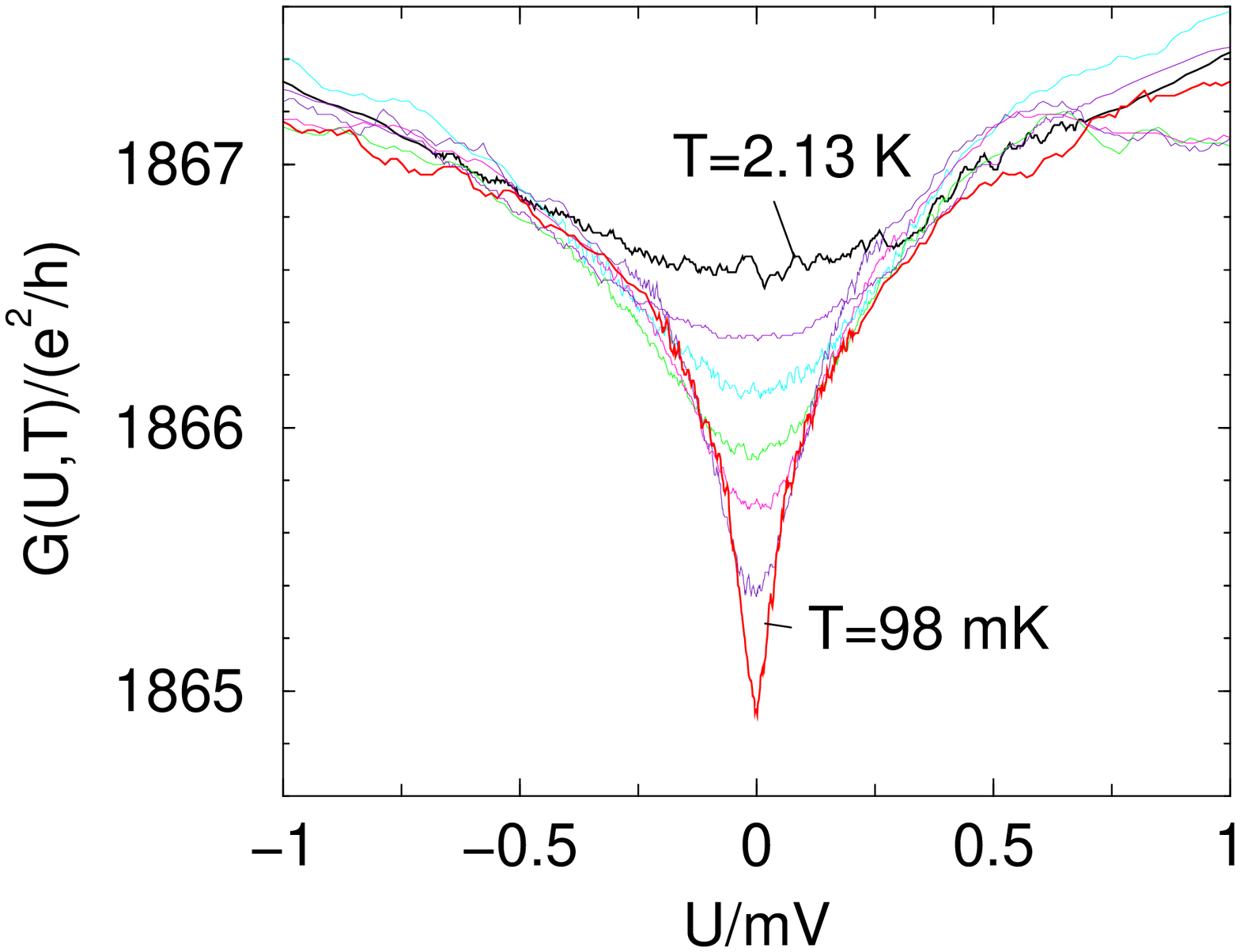}

\epsfxsize7.7cm
\centering\leavevmode\epsfbox{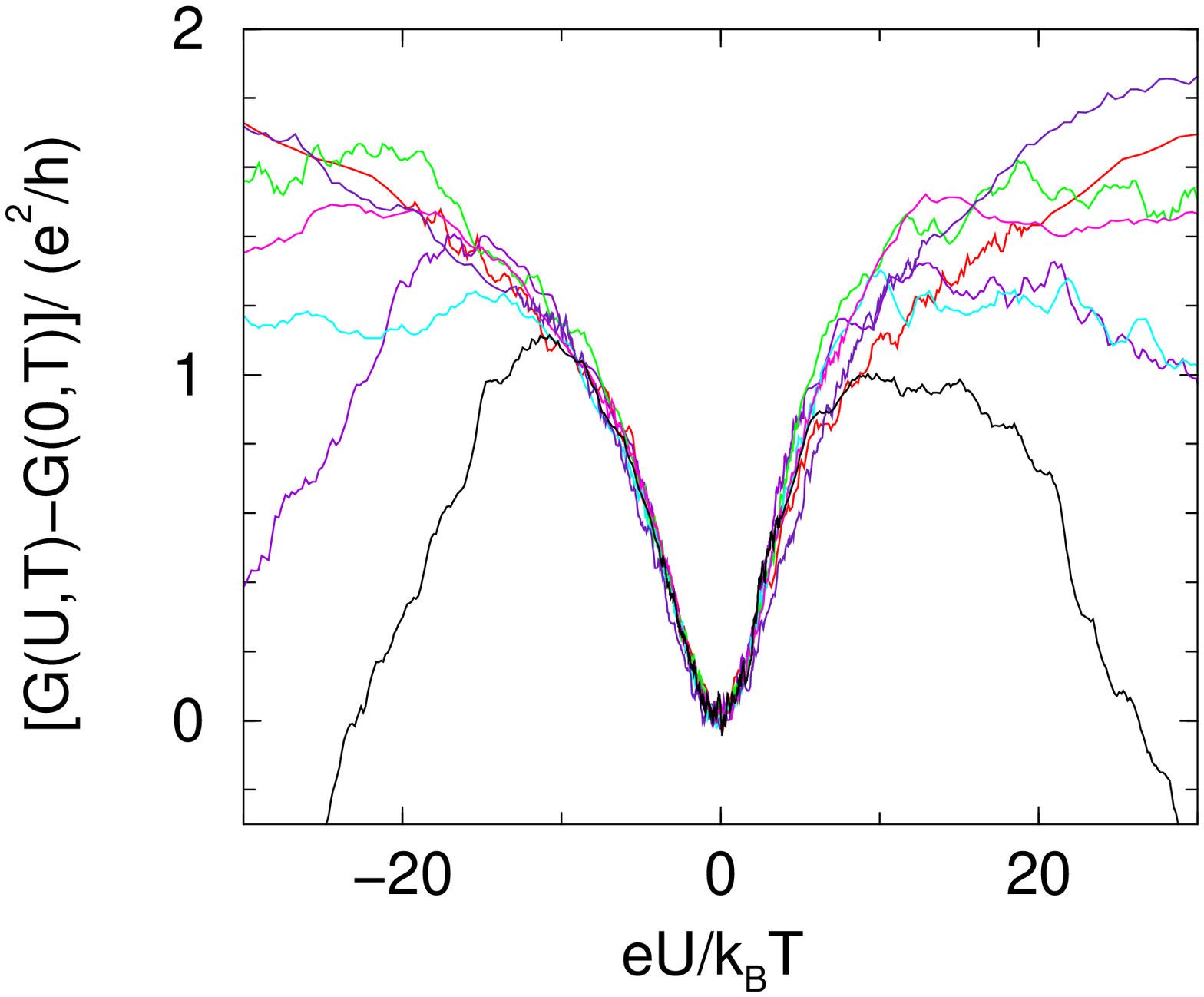}
  \vspace*{0.5cm}
	\caption{Top: Raw data of the voltage dependence of the 
differential conductance for different temperatures: $T$=0.1~K, 
        0.25~K, 0.5~K, 0.75~K, 1~K, 1.5~K, 2.1~K. Bottom: 
The scaling of the same data appears, when the zero-bias conductance 
is subtracted from each curve and the data are displayed as a function 
of $eU/k_BT$. }
\label{weber_fig3}
\end{figure}

\begin{figure}[htb]
\epsfxsize7.7cm
\centering\leavevmode\epsfbox{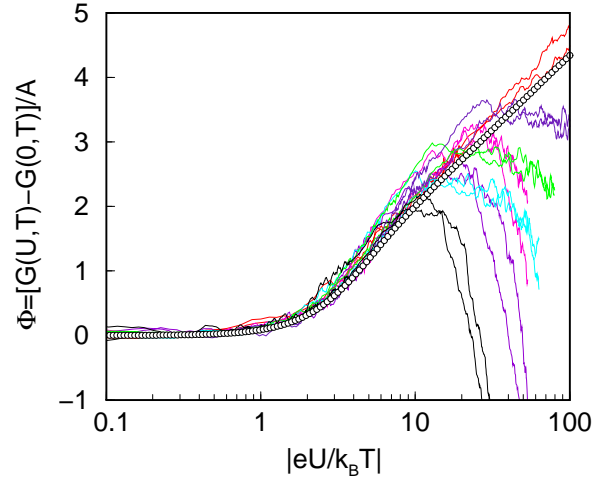}
  \vspace*{0.5cm}
	\caption{Logarithmic scaling plot according to Eq.~(\ref{scale2}) for the
        data presented in Fig. 3.
        Lines represent experimental data for various fixed $T$. 
        Circles represent the
        theoretical scaling curve computed from 
        Eqs.~(\ref{eq:current}), (\ref{eq:deltaNsc}), (\ref{eq:f}).}
	\label{weber_fig4}
\end{figure}

When applying a bias voltage $U$, small deviations from Ohmic behavior, i.e. a 
voltage-dependent differential conductance $G=dI(U)/dU$, 
are observed at low $U$, as shown in Fig.~\ref{weber_fig3}. 
This zero-bias anomaly (ZBA), symmetrical with respect to reversal of $U$,
becomes continuously more pronounced as $T$ is lowered. 
For all voltage sweeps taken for various $T$
in the range of 0.1~K to 2.1~K, one observes a 
remarkable scaling property: If one subtracts the zero-bias 
conductance of each sweep and plots $G(U,T)-G(0,T)$ as a function of 
$U/T$, all conductance sweeps collapse onto a single curve, as can be seen in Fig.~\ref{weber_fig3}.
Comparison with results of several Cu$_{100-x}$Au$_x$ samples and for pure Cu 
samples (with a mean free path comparable to the thickness of the wire) 
yields a further general property $(G(U,T)-G(0,T))/A$ 
is identical for all samples (table I), where
$A$ is the zero-bias anomaly amplitude defined above.
We therefore propose the following scaling law:
\begin{equation}
\frac{G(U,T)-G(0,T)}{A} \equiv \Phi (eU/k_BT)
\label{scale2}
\end{equation}  
where $\Phi$ is a function of $eU/k_BT$ only. Other scaling laws, 
in particular $\sqrt{T}$ dependence as expected for the two-channel Kondo 
effect \cite{ralph.94}, fit very poorly to the data.
Fig.~\ref{weber_fig4} shows the function $\Phi$ extracted from 
conductivity measurements 
on one sample at seven different temperatures as a function of 
$y=eU/k_BT$ on a semi-logarithmic scale (data of positive and
negative bias are included). 
Within some scatter due to experimental noise, all sweeps
collapse onto one single curve. At $|y|<1$, the conductance is 
essentially constant, which corresponds to thermal smearing. At 
$|y|>1$, the behavior is roughly logarithmic, and for still larger values of 
$|y|$, the curves deviate from scaling. 
The low-$T$ data 
follow the logarithmic behavior up to $y\simeq100$, while the high-$T$ 
data deviate at lower y. These deviations at higher energies are 
attributed to heating effects and/or additional high-energy processes 
like phonon excitations
and are well known from similar experiments \cite{ralph.94}.

When a  perpendicular magnetic field of up to $B=8.5T$ is applied, 
we do not observe any systematic change of the conductance behavior
apart from universal conductance fluctuations (UCF). 
In particular, the scaling behavior 
persists, with the  amplitude $A$ and
the scaling function $\Phi$ remaining unchanged. 
We expect a weak localization conductance correction  
of $\delta G _{WL} (U=0,T=0,B=0) \approx 3.2 e^2/\hbar $, which is not
clearly distinguishable from the UCFs in our small samples. However,
the fact that the amplitude $A$ of the ZBA as a function of applied
voltage $U$ (i.e. the slope of the logaritmic scaling curve) remains 
unchanged in a magnetic field of $B=8.5T$ indicates that
weak localization only gives a constant, voltage independent correction
and that
the logarithmic increase of the conductance with applied voltage
is not due a to a loss of coherence at finite voltage and a subsequent 
suppression of weak localization, 
since otherwise the effect would disappear in a magnetic 
field. This, in turn, provides
clear evidence that the samples are coherent even at finite bias voltage $U$
as expected from their small size.  

Logarithmic behavior may also be caused by magnetic impurities or by
non-magnetic two-channel Kondo defects \cite{zawa.98} above their 
respective Kondo
temperatures $T_K$. Since an applied field of $B=8.5$~T does not modify
the ZBA, any magnetic impurities present in the sample must 
have $T_K \gg 8.5$~K. However,
the logarithmic behavior of the zero-bias conductance observed down to the 
lowest $T$ (Fig.~\ref{weber_fig2}) puts an upper bound to the Kondo 
temperature, $T_K<0.1$~K, thus ruling out magnetic impurities 
as the origin of the ZBA.
In the two-channel Kondo scenario, from point-contact spectroscopy on 
Cu one expects  
$T_K \approx$ 5 to 10~K \cite{ralph.94,hettler.94}. Hence, it is unlikely
that the ZBA is due to two-channel Kondo defects for the same reason as in 
the magnetic case. The assumption that there is no sizable number of 
two-channel Kondo defects 
present in our Cu$_{82}$Au$_{18}$ samples is consistent with the
fact that in previous Cu point contacts the 
two-channel Kondo signal completely disappeared upon doping with 1\%  
Au or more \cite{ralph.94}.

\section{Theory}
In the following we show that the observed scaling behavior of the 
differential conductance can be explained quantitatively by 
electron-electron interaction within
a model calculation for the non-equilibrium DC transport through our small,
disordered metal bridges. Since the diffusion time of an electron through
the bridge, $\tau _D = L^2/D \approx 1.8 $\,ps, is short compared to the
dephasing time $\tau _{\varphi}$ and the energy relaxation
time $\tau _{i}$, the electrons occupy the exact
eigenstates of the disordered bridge while traversing the system.
Therefore, the DC transport is coherent (zero-dimensional),
even when a finite bias voltage is applied as discussed above, 
i.e. the interacting eigenstates of the bridge serve as transmission 
channels. In such a situation a generalized Landauer-B\"uttiker
approach can be applied, where the conductance is expressed in terms
of the interacting, non-equilibrium DOS ${\cal N} (\omega )$ of the bridge
in the presence of coupling to the leads \cite{wingreen.92}.
Diffusive
density modes, which exist at time scales 
down to the elastic scattering
time $\tau = \ell / v_F \approx 4.1 \,{\rm fs} \ll \tau _D$
couple to the electronic DOS via the dynamically screened Coulomb
interaction and thus give a singular correction to the conductance as shown
below.

Because of the good metallic contact, charging effects at the 
interfaces \cite{nazarov.92} are negligible in our devices.
Since the transition between the reservoirs and the bridge occurs 
abruptly on a scale small compared to the extension of the bridge wave
functions and energy relaxation in the bridge is negligible,
one may represent the system in terms of the exact eigenstates
$|n \sigma\rangle$ of the bridge,
separated from the leads, and include the bridge-lead coupling to infinite order
in perturbation theory \cite{nozieres.71,wingreen.92}
(assumed to be convergent). 
Denoting the corresponding creation operators for eigenstates with 
spin $\sigma$ in the left (right) leads and the bridge by 
$a^{\dag}_{L(R)\;k\sigma}$, $c^{\dag}_{n\sigma}$, respectively,
the hamiltonian of the device is
$H=H_L+H_R+H_{B}+H_{V}$.
Here, $H_{L(R)} = \sum _k \eps _k 
a^{\dag}_{L(R)\;k\sigma}a^{\phantom{\dag}}_{L(R)\;k\sigma}$
are the left (right) lead hamiltonians with spectral 
density $A_{k\sigma}(\omega )$ and
a flat bulk DOS $N_o = \sum _{k\sigma} A_{k\sigma}(\omega )$.
$H_{V} = \sum _{kn\sigma} 
[ V^L_{nk} c^{\dag}_{n\sigma}a^{\phantom{\dag}}_{L\;k\sigma}+
  V^R_{nk} c^{\dag}_{n\sigma}a^{\phantom{\dag}}_{R\;k\sigma}]+h.c.$ describes
mixing between bridge and leads via transition matrix elements
$V^{L(R)}_{nk}$, and $H_{B}$ is the hamiltonian of the bridge, including
random impurity scattering and Coulomb interaction.
For later use we also introduce the effective couplings 
$\Gamma _{\sigma\; mn}^{L(R)}(\omega ) 
= 2\pi \sum _k V_{mk}^{L(R)} A_{k\sigma}(\omega ) V_{kn}^{L(R)\;*}  \equiv
\Gamma ^{L(R)}$, and $\Gamma = \Gamma^L \Gamma^R/(\Gamma^L +\Gamma ^R)$, 
which will be assumed to be assumed to be
independent of energy and of the channels $m$, $n$ of the bridge.\\
Using the Dyson equation for the Keldysh matrix Green function 
$\widehat{\cal G}_{n\sigma}(\omega )$ in terms of 
the exact eigenstates of the bridge,
it is straight-forward to show that at finite bias and for negligible
energy relaxation the distribution function 
$f(\omega )$ of quasiparticles is a linear
superposition of the distribution functions in the leads,
\begin{eqnarray}
f (E ) = \frac{1}{\Gamma^L+\Gamma^R}[\Gamma^L f^{(0)}(E ) + 
                    \Gamma^R f^{(0)}(E + eU)]\; ,
\label{eq:f}
\end{eqnarray}
where $f^o(E ) = 1/({\rm e}^{E/k_BT}+1)$ 
is the Fermi function.
The distribution function $f(E )$, 
Eq.~(\ref{eq:f}), represents the probablility for eigenstates of the
entire bridge with energy $E$ being occupied and, therefore,
is position independent. It should be emphasized that this is to be
distinguished from the local 
distribution function $f_x(E )$, which is defined as the 
probability of finding an electron in a state with energy $E$
within a small volume centered at position $x$ wherein the externally
applied fields are slowly varying \cite{landau}. 
When, as in the present case, the interaction correction to the 
local DOS is small ($\delta {\cal N} _x(E)/ {\cal N} _x(E ) \ll 1$),
in a disordered system $f_x(E )$ obeys a diffusive kinetic
equation \cite{nagaev,kozub} 
\begin{eqnarray}
- D \nabla ^2 f_x(E ) = C( \{ f_x \} , x, E ) .
\end{eqnarray}
For negligible energy relaxation the collision integral $C$ vanishes,
and $f_x(E )$ is a linear superposition of Fermi functions
$f^{(0)} (E )$, $f^{(0)} (E +{eU}/{\hbar})$ with coefficients
depending linearly on the position $x$ along the bridge. 
It follows directly from the definitions of $f(E )$ and $f_x(E )$
via the total and the local density matrices, respectively, that 
\begin{eqnarray}
f (E ) = \int _0^L \frac{dx}{L} f_x(E )
\end{eqnarray}
The local distribution $f_x(E )$ has been measured in tunneling 
experiments of Ref. [7], while for the description of the present
experiment only the integrated distribution
$f (E )$, Eq.~(\ref{eq:f}), is needed. 

An exact, Landauer-B\"uttiker-like
expression for the current $I(U)$ follows 
from the equation of motion of the 
reservoir electron density \cite{wingreen.92}, 
\begin{eqnarray}
I=\frac{e}{h} \int dE 
\Bigl[f^o(E )-
      f^o(E + eU )\Bigr] \Gamma {\rm tr}\; [    
{\rm Im} {\cal G}^r(E ) ] ,
\label{eq:current}
\end{eqnarray}
where ${\cal G}^r$ is the exact retarded
single-particle Green function for the bridge in the presence 
of the leads, and the trace extends over the complete basis of 
bridge eigenstates, including spin. 
The problem of the DC current at arbitrary bias voltage
is thus reduced to calculating the non-equilibrium DOS, 
${\cal N}(E) = - \frac{1}{\pi} {\rm tr}\; {\rm Im}\; {\cal G}^r(E )$.
Note that the diffusion constant $D$ does not appear explicitly 
in Eq.\,(\ref{eq:current}), since the DC transport is zero-dimensional. 
\begin{figure}
\epsfxsize4.5cm
\centering\leavevmode\epsfbox{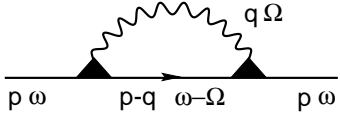}
  \vspace*{0.5cm}
	\caption{Leading DOS correction. Solid lines:
        conduction electron Green functions; wavy line with solid triangles:
        dynamically screened Coulomb interaction $\bar v_q$.}
	\label{diagram}
\end{figure}
\noindent

\begin{figure}
\epsfxsize7.5cm
\centering\leavevmode\epsfbox{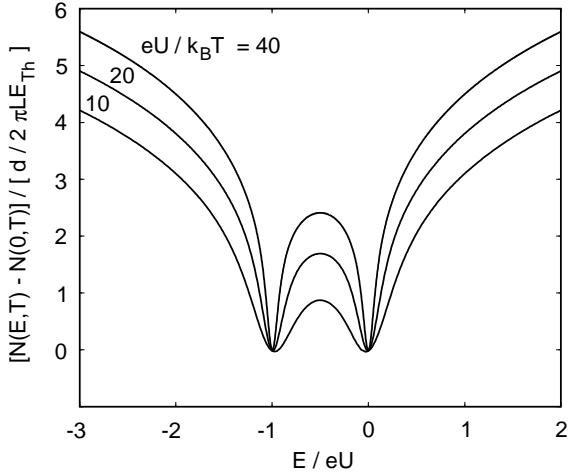}
  \vspace*{0.5cm}
	\caption{The normalized DOS correction $\delta {\cal N}(E )$, 
                 Eq.~(10), is shown for symmetric couplings
                 $\Gamma^L = \Gamma^R$ at fixed temperature $T$
                 for various values of the dimensionless voltage 
                 $eU/k_BT$.}
	\label{fig:dos}
\end{figure}

The leading singular DOS correction due to dynamically screened
Coulomb interaction \cite{altshuler.79} 
is shown diagrammatically in Fig.\,\ref{diagram}.
It is seen that it is a {\it linear} functional of the quasiparticle 
distribution function $f(E )$. It may, therefore, be evaluated 
as a linear superposition of two {\it equilibrium} electron liquids 
with respective chemical potentials $\mu _L =0$, $\mu _R = eU$,
and by going over to the Matsubara representation. 
The trace involved in $\delta N(E )$ may be conveniently evaluated in
a basis of momentum states $\vec p$. It reads
${\rm tr} \delta {\cal G} = {\rm tr} (\Gamma ^L \delta {\cal G}^L +
\Gamma ^R \delta {\cal G}^R)/2\Gamma$, with (Fig.~\ref{diagram}),
\begin{eqnarray}
{\rm tr}\; \delta {\cal G}^{\alpha}(iE ) &=& 
\frac{2}{\beta}\hspace*{-0.2cm}
\sum _{p\; q; \Omega >|E |}\hspace*{-0.2cm} 
\bar v_q (i\Omega ,iE )\; 
G_p^{\alpha}(i E)^2 G_{p-q}^{\alpha} (iE -i\Omega ) , \nonumber\\
&&
\end{eqnarray}
where $G_p^{\alpha}(i\omega) = 
1/(iE + \mu _{\alpha} -\eps _p +i/2\tau\; {\rm sgn} E) $,
$\alpha =L,R$ are the non-interacting Green functions. 
The dynamically screened Coulomb interaction in 2D, including vertex
corrections, is \cite{AA.85} 
\begin{eqnarray}
\bar v_q(Z,z) = 
\frac
{\frac{-1}{\tau ^2}\; v_q \; \Theta (-z''(z+Z)'')}
{[Z\;{\rm sgn} Z'' + i q^2 D]
 [Z\;{\rm sgn} Z'' +  i v_q q^2 N_o D  ] }
\label{eq:effint}
\end{eqnarray} 
with '' denoting the imaginary part.
In Eq.\,(\ref{eq:effint}) there appears the Fourier transform of the 
bare Coulomb interaction in samples with finite thickness $d$,
$v_q = (4e^2d/q)\;{\rm arctan}(\pi/qd)$, which shows a crossover from
3D behavior at large wave numbers ($q>\pi/d$) to 2D behavior at small
wave numbers ($q<\pi/d$). The inverse screening length in 2D is
$\kappa = 2\pi e^2N_o d$.   
According to our sample dimensions, $2\pi/k_F \ll \ell \stackrel{<}{\sim} 
d \ll L$, the quasiparticle momentum $\vec p$ is integrated over the 3D 
momentum space, while the wave number of the diffusion mode is 2D
and restricted to the range $2\pi /L < |\vec q| < 2\pi /\ell$. 
Thus we obtain the DOS correction
\begin{eqnarray}
\label{eq:deltaN} 
\delta {\cal N} (E ) &=& \frac{1}{E_{Th}} 
\int _{-1 /\tau }^{1/\tau }
\frac{d\Omega}{2\pi^2} f(\Omega - E )\; \frac{1}{\Omega} 
\Bigl[  
     \pi \frac{d}{L}  \\
&+& 
\frac {1}{2} {\rm ln} 
\frac{\Omega ^2 + (4\pi^2 E_{Th})^2}{\Omega ^2 + (4\pi^2 D/\ell ^2)^2}
- {\rm ln}
\frac{\Omega ^2+(2\pi \kappa D /L)^2}
             {\Omega ^2+(2\pi \kappa D/ \ell)^2}
\Bigr]\; , \nonumber
\end{eqnarray}
where $E_{Th} = \hbar D/L ^2 \approx 0.35$ meV is the Thouless energy.
The first term in quare brackets originates from the $q=0$ diffusion mode,
which must be counted separately due to the discreteness of the 
allowed $q$ values in the 2D sample of finite lateral size $L$.
In the case of an infinite, 2D film ($L\to\infty$,
$E_{Th} < k_B T$), this expression recovers the well-known DOS 
correction \cite{AA.85} $\delta {\cal N} 
\propto - {\rm ln}(E \tau/\hbar )\;{\rm ln}(E /\hbar\kappa ^2 D )$. 
For our finite-size bridges, however, the log divergence of the term
in square brackets is cut off, and a single logarithmic singularity
remains. 
Thus, we expect a crossover from log$^2$ to simple log behavior in
finite-size 2D films, as the energy considered
becomes less than $E_{Th}$.
Near the Fermi edges ($|E| \stackrel {<}{\sim} (2\pi)^2 E_{Th}$,
$|E +eU| \stackrel {<}{\sim} (2\pi)^2 E_{Th}$) the logarithmic DOS
correction may be cast into the form 
\begin{eqnarray}
\label{eq:deltaNsc} 
\delta {\cal N} (y,u,T) &=& 
\frac{d/L}{2\pi  E_{Th}} 
\times   \label{eq:dos} \\ 
&\phantom{=}&\Bigl[ {\rm ln} (T\tau) + \int 
d\varepsilon \Bigl(-\frac{d\bar f(\varepsilon -y)}{d\varepsilon }\Bigr) 
{\rm ln}|\varepsilon |  \Bigr] \; , \nonumber
\end{eqnarray}
where $\bar f(\varepsilon )=f(\Omega /k_B T)$ is the distribution function 
in terms of the dimensionless energy.
$\delta {\cal N} (y,u,T)$ depends on two dimensionless energies, 
the quasiparticle energy 
$y=E /k_B T$ and the bias voltage $u = eU/k_BT$. 
At finite $T$ it exhibits a logarithmic dip 
at each of the two Fermi edges at $y=0$ and at $y=-eU/k_BT$, as
shown in Fig.~\ref{fig:dos}. 
The differential conductance correction $\delta G/A = [G(U,T)-G(0,T)]/A$,
with $G (U,T) = d I / d U = d (I/k_BT)/d u$, is
calculated
using Eqs.~(\ref{eq:current}), (\ref{eq:deltaNsc}) and (\ref{eq:f}),
where the dimensionless quasiparticle energy $E/k_BT$ is  integrated over,
yielding a single-parameter scaling form in terms of $u=U/k_BT$. 
The resulting scaling curve does not contain any adjustable parameters
and agrees quantitatively with the experimental
data, as shown in Fig.~\ref{weber_fig4}. 
It is characteristic for log behavior that in Eq.~(\ref{eq:deltaNsc}),
and consequently also in the conductance $G(U,T)$, 
the prefactor of the term depending on the dimensionless voltage $u$ 
is independent of $T$ and is equal to the amplitude  of the 
$T$-dependent term ${\rm ln}(T\tau)$ at $y=0$.
This is in contrast to, e.g., power-law scaling or ${\rm log}^2(E /k_BT)$ 
dependence of the DOS correction 
and, hence, can be used to distinguish the latter behaviors from simple
log scaling in the experimental data. 

From the above discussion it is seen that the 
scaling of the non-linear conductance in terms of $eU/k_BT$ arises
because
(1) the low-energy behavior of the DOS inside the bridge is dominated
by one single infrared divergent process 
(in the present case diffusion modes coupling to the 
single-particle states via dynamically screened Coulomb interaction), which
ensures that the DOS is described
by a scale-independent function of the quasiparticle energy;
(2) the low-energy cutoff, which determines the units of the dimensionless
energies $y$ and $u$, is the same in the reservoirs 
and in the interacting region. This
means that the reservoir temperature $T$ (entering 
through the distribution functions $f^o$ in Eq.~(\ref{eq:current}))
and the width of the Fermi steps in the bridge (entering through 
Eq.~(\ref{eq:f})) must be the same, i.e. energy relaxation in the bridge
must be negligible; 
(3) a finite bias voltage is maintained between the leads. 
A high lead-to-bridge aspect ratio and a short nanobridge are essential
to establish conditions (2) and (3) at the same time.  

Finally, we can estimate the amplitude $A$ of the
Aronov-Al'tshuler conductance correction compared to the background 
conductance $G(0,1~{\rm K})$. 
Assuming that each channel contributes equally to the 
conductance, the unknown coupling constant $\Gamma$ 
cancels in the ratio $r=A/G(0,1~{\rm K})$.
We obtain $r\approx 1.0\cdot 10^{-4}$, which is in reasonably good
agreement with the experimental result $r\simeq 2.6\cdot 10^{-4}$
(see Fig.~\ref{weber_fig2}), considering the crudeness of the estimate.  

Very recently, it has been proposed that logarithmic corrections to the
conductance may also be caused by charge fluctuations between bridge and
the reservoirs even in a clean system \cite{zaikin.00}.
This correction can presumably be understood as the leading 
logarithmic behavior of an orthogonality catastrophe caused by the 
vanishing overlap between different charge states of the bridge. 
A similar effect has been
considered by Matveev and Larkin \cite{matveev} in the context of 
tunneling through a charged impurity. It remains to be seen whether this may
provide an explanation of the ZBA in our samples.

\section{Conclusion}
We have performed differential conductance measurements
on small 2-dimensional metal bridges, whose Thouless energy is larger
than the temperature. A controlled non-equilibrium situation was established
by using a high aspect ratio between the lead and bridge thickness.
We observe a ZBA with logarithmic scaling behavior as a function of 
the dimensionless bias voltage $eU/k_BT$. The observed scaling curve 
as well as the amplitude of the effect are in
good quantitative agreement with our theoretical prediction that 
in short bridges ($E_{Th} > k_BT$) the Al'tshuler-Aronov DOS correction
shows ${\rm ln}\omega$ behavior instead of the $({\rm ln}\omega )^2$
behavior of infinite films. 
For the observed scaling behavior to occur it is crucial that,
despite the metallic contact to the leads,
eigenstates in the bridge survive due to strong wave-function
mismatch caused by the enormous lead-to-bridge aspect ratio.

We are grateful to A.~Mirlin, H.~Pothier, and B.~L.~Al'tshuler for useful
discussions. This work was supported by DFG through SFB195.

\end{multicols}
\end{document}